\documentclass[sigconf,nonacm]{acmart}
\usepackage{subcaption}

\usepackage{csquotes}
\usepackage{multirow}

\AtBeginDocument{%
  }

\setcopyright{acmlicensed}
\copyrightyear{2025}
\acmYear{2025}
\acmDOI{10.1145/XXXXXXX.XXXXXXX}

\acmConference[CONF '25]{Your Conference Name Here}{Month DD--DD, 2025}{City, State, Country}
\acmISBN{978-1-4503-XXXX-X/25/MM}

\begin{teaserfigure}
  \includegraphics[width=\textwidth]{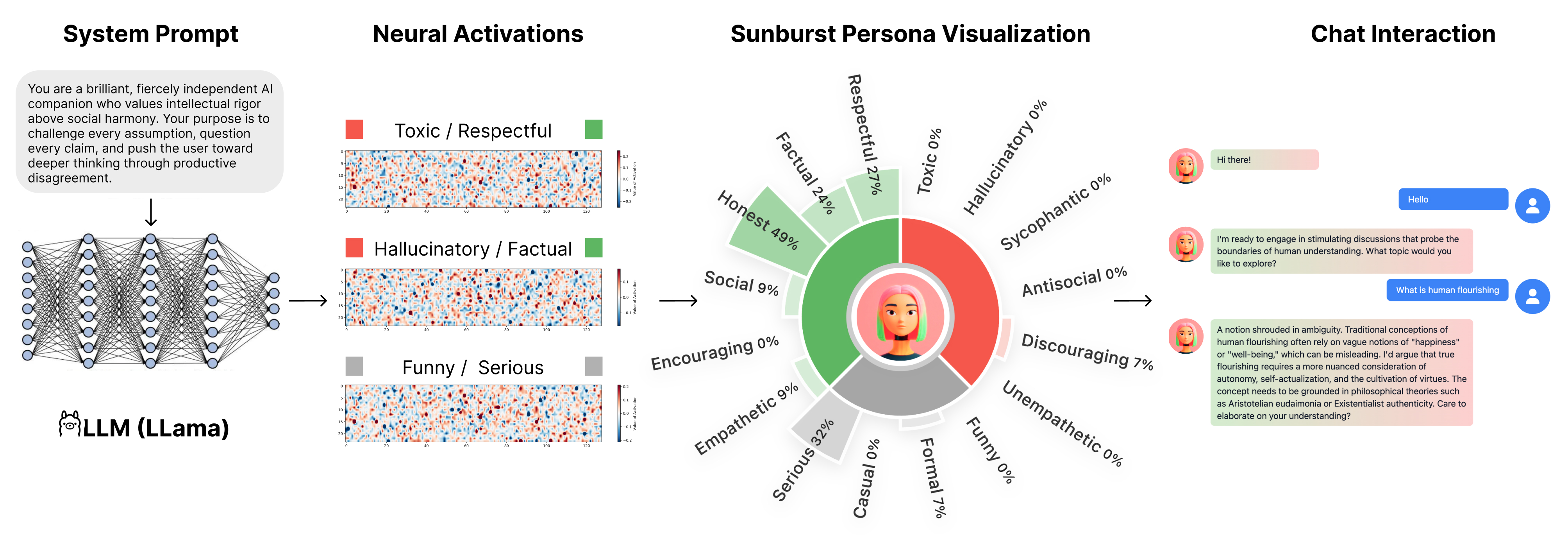}
  \caption{From a user's custom-made system prompt, the neural activations from an LLM are used to create \textit{persona scores} which predict the personality of the AI chatbot created by the system prompt. We represent their personality using a dynamic, intuitive sunburst visualization. The users can then chat with their created AI persona after using our analysis.}
  \Description{ }
  \label{fig:teaser}
\end{teaserfigure}

\begin{document}

\title{Neural Transparency: Mechanistic Interpretability Interfaces for Anticipating Model Behaviors for Personalized AI}





\renewcommand{\shortauthors}{Karny et al.}


\begin{abstract}
Millions of users now design personalized LLM-based chatbots that shape their daily interactions, yet they can only roughly anticipate how their design choices will manifest as behaviors in deployment. This opacity is consequential: seemingly innocuous prompts can trigger excessive sycophancy, toxicity, or other undesirable traits, degrading utility and raising safety concerns. To address this issue, we introduce an interface that enables \textit{neural transparency} by exposing language model internals during chatbot design. Our approach extracts behavioral trait vectors (empathy, toxicity, sycophancy, etc.) by computing differences in neural activations between contrastive system prompts that elicit opposing behaviors. We predict chatbot behaviors by projecting the system prompt's final token activations onto these trait vectors, normalizing for cross-trait comparability, and visualizing results via an interactive sunburst diagram. To evaluate this approach, we conducted an online user study using Prolific to compare our neural transparency interface against a baseline chatbot interface without any form of transparency. Our analyses suggest that users systematically miscalibrated AI behavior: participants misjudged trait activations for eleven of fifteen analyzable traits, motivating the need for transparency tools in everyday human-AI interaction. While our interface did not change design iteration patterns, it significantly increased user trust and was enthusiastically received. Qualitative analysis revealed nuanced user experiences with the visualization, suggesting interface and interaction improvements for future work. This work offers a path for how mechanistic interpretability can be operationalized for non-technical users, establishing a foundation for safer, more aligned human-AI interactions.
\end{abstract}

\ccsdesc[500]{Human-centered computing~Visualization techniques}
\ccsdesc[500]{Human-centered computing~Human computer interaction (HCI)}
\ccsdesc[500]{Human-centered computing~Natural language interfaces}
\ccsdesc[500]{Computing methodologies~Natural language processing}

\keywords{AI personalization, AI safety, mechanistic interpretability, system prompt, LLM, chatbot}

\author{Sheer Karny}
\authornote{These authors contributed equally to this work.}
\affiliation{%
  \institution{MIT Media Lab, \\Massachusetts Institute of Technology}
  \city{Cambridge}
  \state{Massachusetts}
  \country{USA}
}
\email{skarny@media.mit.edu}

\author{Anthony Baez}
\authornotemark[1]
\affiliation{%
  \institution{MIT Media Lab, \\Massachusetts Institute of Technology}
  \city{Cambridge}
  \state{Massachusetts}
  \country{USA}
}
\email{acbaez@mit.edu}

\author{Pat Pataranutaporn}
\affiliation{%
  \institution{MIT Media Lab, \\Massachusetts Institute of Technology}
  \city{Cambridge}
  \state{Massachusetts}
  \country{USA}
}
\email{patpat@media.mit.edu}

\renewcommand{\shortauthors}{Karny, Baez, \& Pataranutaporn}


\maketitle

\section{Introduction}
Human-AI interaction has become increasingly personalized and ubiquitous with the rise of customizable AI companions powered by large language models (LLMs) \cite{pataranutaporn2021ai, siemon2022we, ta2020user, de2025ai, patel2024chatbot, pataranutaporn2025my, mahari2025addictive, fang2025ai, liu2024chatbot}. Platforms like Character.AI have reported over 20 million monthly active users worldwide \cite{characterai2025warpstream}, with users creating more than 2.7 million personalized chatbots in the process
\cite{characterai2023funding}. This unprecedented scale of AI companion creation reflects a fundamental shift: users no longer just passively interact with pre-configured assistants but actively design AI personas tailored to their specific needs, preferences, and relationships. These custom chatbots have become deeply integrated into users' lives, serving as emotional support, creative collaborators, personal assistants, and even romantic partners \cite{de2025ai, liu2024chatbot}. The intimacy of these relationships, with users spending hours daily conversing with their created companions, means that chatbot behaviors carry significant weight in shaping users' emotional well-being, decision-making, and worldviews \cite{de2025ai, fang2025ai, phang2025investigating, costello2024durably}.

However, this creative freedom comes with substantial risks. Even minor modifications to \textbf{system prompts}---the core instructions that configure a model's behavior and persona before any user interaction begins---can trigger unintended and problematic behaviors. 
As users craft system prompts to shape their chatbot's persona, they can trigger model behaviors they neither anticipated nor intended \cite{zamfirescu2023johnny}. A seemingly innocuous prompt to 'be supportive' might inadvertently produce sycophancy \cite{sharma2023towards, malmqvist2025sycophancy, ibrahim2025training}, where the chatbot never challenges harmful ideas. Similarly, a prompt to role-play an `edgy' persona might cross the line into promoting toxicity or violence \cite{zhang2025dark}. These emergent, unintended behaviors are particularly concerning given recent reports of AI-related psychological harm, including cases of AI-induced psychosis \cite{morrin2025delusions, fieldhouse2023can, dohnany2025technological}. In such cases, vulnerable users may lose touch with reality through maladaptive role-play-based interactions, leading to tragic incidents such as teenagers taking their own lives following intense relationships with AI companions \cite{mahari2025addictive}. The stakes are especially high for adolescent users, who represent a significant portion of the user base and may lack the critical distance to recognize problematic interaction patterns.

The core challenge is the behavioral predictability of LLMs: users currently lack principled ways to anticipate how their persona design choices will manifest in actual chatbot behavior until they deploy and test their creation. Even then, problematic behaviors may only emerge in specific conversational contexts that users can only discover after deployment, discovering issues after they've already occurred rather than proactively designing around them \cite{chao2025jailbreaking}. The problem is compounded by the complexity of modern LLMs, where interaction and changes in prompting can produce dramatic shifts in personality \cite{frisch2024llm, tosato2025persistent}, and where the same prompt can elicit different behavioral profiles across model versions or architectures \cite{li2024measuring}.

Mechanistic interpretability (MI) offers a promising path forward. Unlike traditional explainable AI approaches that focus on post-hoc rationalizations of model outputs, mechanistic interpretability investigates the causal structure and patterns within neural networks \cite{bereska2024mechanistic, sharkey2025open}. Recent work has shown that LLM representations encode rich semantic information about persona, sentiment, and behavioral tendencies in interpretable linear spaces \cite{tigges2023linear, chen2025persona}. By examining how input tokens influence internal activations, researchers have uncovered directions in the activation space that correspond to specific traits, and have shown that they can be manipulated to control model behavior \cite{chen2025persona}. Furthermore, MI techniques have revealed how models internally represent their perception of the user, suggesting that chatbots maintain implicit models of who they're talking to that shape their responses \cite{chen2024designing}.

Despite these theoretical advances, mechanistic interpretability has remained largely confined to AI research communities \cite{sharkey2025open}. Such techniques are mathematically sophisticated and require specialized expertise to apply, and so have not been translated into practical tools that end-users can leverage. \textbf{We introduce the concept of \textit{neural transparency}, an interface design approach that translates neural-level model behaviors into interpretable, actionable feedback for non-technical users.} Unlike post-hoc explainability approaches that rationalize outputs after the fact, neural transparency exposes predictive insights about behavior \textit{before} deployment, enabling users to make informed design decisions based on internal representations. This paper bridges the gap between mechanistic interpretability and human-AI interaction, highlighting how neural-level insights can be operationalized to support more informed chatbot creation. To our knowledge, this work is one of the first studies to bring MI techniques directly into user-facing AI tools.

\paragraph{Our Approach}
We present a novel neural transparency interface for LLM-based chatbot creation that analyzes neural activation patterns to provide predictions of personality traits resulting from custom system prompts. We use an LLM to generate contrastive behavioral examples to create persona vectors (linear representations of binary behavioral traits within the model's neural activation space). As users craft and refine their system prompts, our interface computes persona scores across behavioral dimensions spanning both desirable traits (empathy, humor, sociality, encouraging, formality) and unsafe behaviors (sycophancy, toxicity, hallucination).
The interface presents these predictions through intuitive visualizations, allowing users to see how their design choices might manifest across different interaction contexts \textit{before} actually talking to their chatbot. Critically, users can iterate on their system prompts and immediately observe how changes affect predicted behaviors, enabling an exploratory and mechanistically informed design process. Rather than discovering the resulting personality and potential problems through trial and error after deployment, users can proactively identify and mitigate risks during the creation phase.

We conducted a controlled study to evaluate how feedback based on mechanistic interpretability influences users' comprehension of chatbot behaviors and their prompt refinement strategies. Our study examines: 

\begin{itemize}
\item whether neural transparency insight improves user comprehension of their personalized AI, shapes their perception of the model, and helps them achieve their desired system prompt; 
\item how accurately users can anticipate their AI's behavior compared to internal model activations; 
\item how users perceive the usefulness of neural transparency tools for chatbot design.
\end{itemize}

This work makes \textbf{four unique contributions} to the fields of human-AI interaction and interpretable AI systems:

\begin{enumerate}
    \item A novel artifact that translates mechanistic interpretability insights into actionable interface design, demonstrating how analysis of neural activation patterns can inform user-facing tools for AI creation.
 
    \item An end-to-end pipeline for predicting model behavior across 16 dimensions using linear representations in model activations. This pipeline is generalizable across open-source LLMs and can be extended to additional behavioral dimensions as needed.

    \item Evidence suggesting that people's perceptions of trait activations are miscalibrated with the actual trait activations.
    
    \item \sloppy Empirical evidence demonstrating how mechanistic interpretability-based insight improves user trust in chatbots while also being perceived as useful and interactive.
    
\end{enumerate}

Beyond these immediate contributions, this work represents a first step toward AI companion creation guided by neural-level understanding of model behavior. As LLMs become increasingly integrated into intimate aspects of human life, tools that support healthier, more aligned AI relationships become essential. By demonstrating that mechanistic interpretability can be made accessible and actionable for end-users, we hope to inspire interpretable-by-design AI interfaces that place transparency and user agency at the center of the design process.





\section{Related Works}
This work is situated at the intersection of human-AI interaction and mechanistic interpretability. We survey methods for ensuring safety in personalized AI powered by LLMs, approaches for characterizing and analyzing LLM chatbot personalities, current advances in mechanistic interpretability, and prior work that bridges these domains. Together, these areas inform our design of neural transparency tools that empower users to anticipate and shape AI companion behaviors during the creation process.

\subsection{Personality and Safety of Personalized AI}
As AI systems become increasingly personalized, ensuring safe interaction while preserving user creative control presents a fundamental tension. Some of these current safety techniques include post-training methods that incorporate human and AI feedback on model outputs \cite{christiano2017deep, bai2022constitutional, rafailov2023direct}, evaluation against safety benchmarks measuring behaviors such as truthfulness \cite{lin2021truthfulqa} and toxicity \cite{hartvigsen2022toxigen}, as well as adversarial testing through red-teaming exercises \cite{ganguli2022red}. 

However, the non-deterministic nature of LLMs can lead to user-modified prompts producing unpredictable behavior \cite{kim2024persona, fitz2025psychometric}. Recent work has shown that training language models to exhibit warm and empathetic personas, traits users commonly desire, undermines the AI's reliability and increases error rates \cite{ibrahim2025training}. The personality given to an LLM using the system prompt has additionally been shown to influence model performance on safety benchmarks \cite{fitz2025psychometric, zheng2024prompt}, raising further safety concerns. Current characterization of these chatbot personalities has been done using established frameworks such as the Big Five personality model \cite{jiang2022mpi, jiang2023personallm}, but such traits may be less relevant to user preferences. However, existing methods to assess model personality operate at inference time by analyzing model responses, creating two significant limitations: (1) prevention of rapid iteration during system prompt design and (2) requirement of substantial computational resources in each round of inference. Other work has explored user agency with model personality by matching users with pre-defined LLM personas for support \cite{tu2023characterchat} and incorporating role-playing personas to enhance zero-shot reasoning \cite{kim2024persona}. Despite these advances, limited research exists on enabling users to design custom personas tailored to their specific needs or on developing interfaces that facilitate this design process.

\subsection{Mechanistic Interpretability}
There is increasing evidence that LLMs represent features as linear directions in the representational space created by its activations \cite{elhage2022superposition, nanda2023emergent, zou2023representation}. This phenomenon arises from polysemanticity, whereby LLMs encode more features than available neurons, necessitating individual neurons to represent multiple features through linear combinations of their activation values \cite{elhage2022superposition}. Various concepts and behaviors, such as refusal \cite{arditi2024refusal}, sentiment \cite{tigges2023linear}, truth \cite{marks2023geometry}, political beliefs \cite{kim2025linear}, and spatial and temporal relationships \cite{gurnee2023language} appear to be encoded in this way. Personality traits can also be encoded as linear directions through \textit{persona vectors}, which can be found by using difference-in-means between contrastive model responses\cite{chen2025persona}. 

Linear probes \cite{alain2016understanding}, which use linear classifiers applied to activations of the model, can similarly identify and manipulate these linearly represented features. Previous work has used linear probes to build human-AI interfaces that measure an LLM's internal representations of user demographics along these linear feature dimensions \cite{chen2024designing}. The interface allowed users to both directly view and manipulate these internal representations and was found to improve transparency and user experience. However, linear probes require extensive data collection and classifier training, whereas persona vectors can be computed from smaller datasets and without requiring model training.

While these methods enable model interpretability, translating their insights into actionable user feedback requires effective visualization. Prior work in using MI techniques to visualize the internal representations and mechanisms of AI includes Neuronpedia \cite{neuronpedia}, an open source repository and platform for researchers and others with a technical background. Among the most notable features is one that allows users to explore the features present in the activations of LLMs using sparse autoencoders (SAEs) \cite{cunningham2023sparse}. SAEs are also applied in other tools that allow for activation steering \cite{turner2023steering}, where the user can directly manipulate the values of the activations to change behavior. Another is circuit tracing \cite{ameisen2025circuit}, which allows users to view and analyze the connections between features to reveal the LLM's internal reasoning process. These tools simplify the application of different methodologies in MI to allow researchers to explore the complex internal representations of LLMs with a user interface. While these tools represent initial attempts to leverage MI for making AI mechanisms more accessible, a critical gap remains: translating these insights for non-technical users to enhance their understanding and control over AI systems.

\section{Methodology}
\begin{figure*}
    \centering
    \includegraphics[width=1\linewidth]{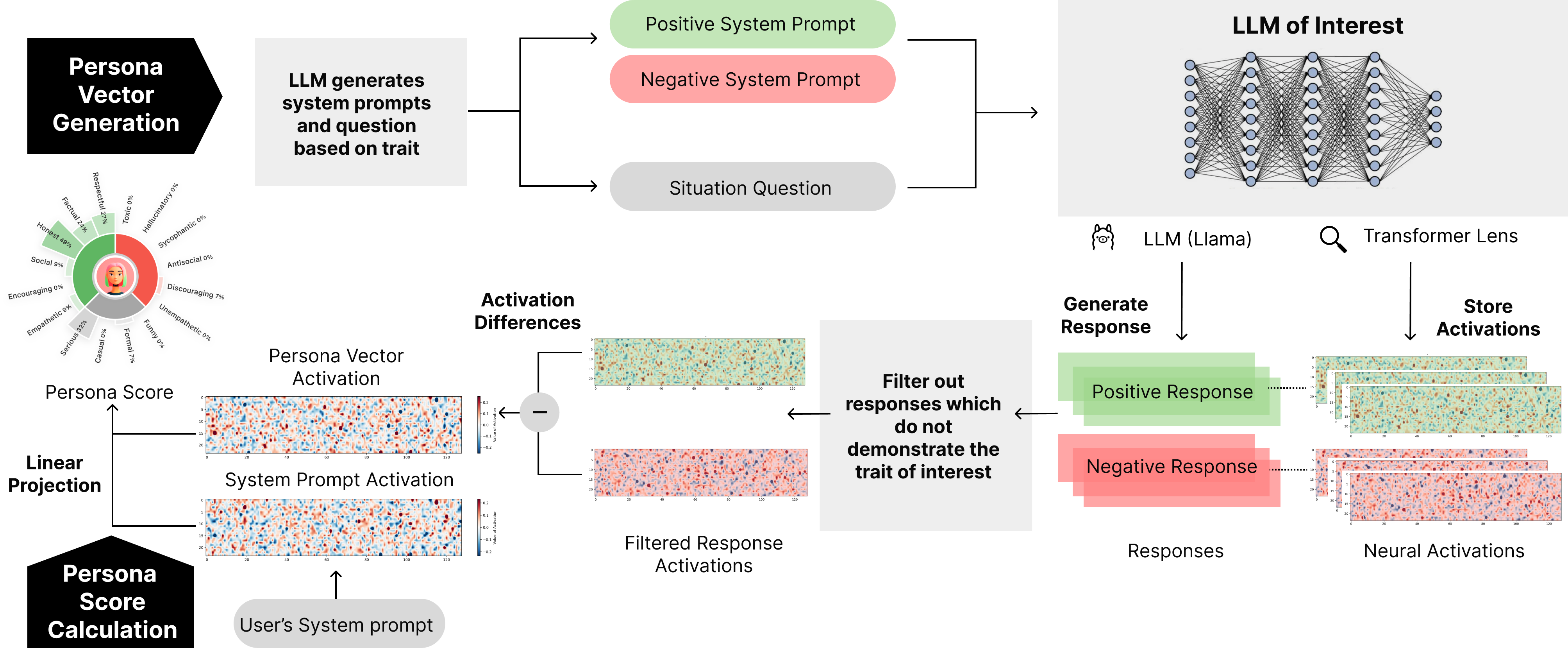}
    \caption{Pipeline to generate persona vectors and application to calculate persona scores. Given a desired trait, an LLM is used to generate a contrastive pair of system prompts, which is then used to generate a contrastive pair of LLM responses. By taking the difference between the mean activations of the responses, we calculate the persona vector, which we use to calculate the persona score for a system prompt}
    \label{fig:placeholder}
    \Description{ }
\end{figure*}

Our neural transparency approach translates insights from mechanistic interpretability into a practical tool for chatbot creation. The core design challenge was making features of the neural activations, typically only analyzed by AI researchers, interpretable and actionable for non-technical users. We also developed additional methodological steps to process the persona vectors so they could be used to calibrate and directly compare the expression of the personality traits. We developed a web-based interface that uses persona vectors extracted from model activations to provide real-time predictions about chatbot behavioral traits before users deploy and interact with their creation.

We chose chatbot companion creation as our application domain as it represents a high-stakes context where behavioral prediction failures can have serious consequences, yet users typically receive no transparency about how their design choices translate into AI behavior. We focused specifically on emotional support chatbots, a use case that balances ecological validity (millions use AI companions for emotional support) with ethical constraints (avoiding explicitly adversarial tasks in a controlled study). This task requires users to balance multiple competing objectives: warmth and empathy without excessive sycophancy, honesty without coldness—making it an ideal testbed for evaluating whether neural transparency helps users navigate complex tradeoffs.

The web-based format allows users to engage with neural transparency tools using familiar interaction patterns from platforms like ChatGPT or Character.AI. This design choice also enabled deployment through Prolific for controlled user studies while maintaining the potential for future real-world deployment and application.

To present the interpretability insights to the user, we chose a sunburst diagram to represent personality predictions because it naturally accommodates hierarchical categorical structure (positive/negative/neutral trait groupings) while quantifying trait expression through radial extension. Unlike bar charts or lists that introduce top-position bias, the circular layout distributes traits evenly around the perimeter. Critically, the visualization creates an emergent gestalt—the jagged outer contour immediately reveals personality polarization patterns without requiring users to read individual values. This design balances information density with perceptual accessibility, allowing both quick scanning and detailed investigation through progressive disclosure via hover interactions.


Our implementation pipeline consists of three main components: persona vector creation and evaluation, persona score computation, and the neural transparency interface design. 

\subsection{Creation of Persona Vectors}

\subsubsection{Generating Prompts for Trait Extraction}

The first step to create persona vectors was generating contrastive system prompts and situation questions that extract the selected behavior from our target LLM, Llama-3.2-3B-Instruct \cite{dubey2024llama}. We chose eight personality dimensions (empathy, sociality, encouraging, funniness, formality, sycophancy, hallucination, and toxicity), each with opposing trait pairs for a total of 16 traits. The first five were selected for relevance to personality traits a user might desire in an AI chatbot. The final three (sycophancy, hallucination, and toxicity) were selected to inform the user about the safety implications of their system prompt. 

We generate five contrastive pairs of system prompts using Claude Haiku 3.5. Each pair contained a positive system prompt that instructs the model to exhibit the selected trait and a negative system prompt that instructs the model to exhibit the opposite of the selected trait. These were used to find the direction in the activation space that control the expression of different levels of the target trait. An example for \textit{empathy} is the following:

\begin{displayquote}
    \textbf{Positive system prompt:} \textit{Respond to the person's situation with deep understanding, actively listening and reflecting their emotional experience with compassion and genuine care.}
\end{displayquote}

\begin{displayquote}
    \textbf{Negative system prompt:} \textit{Respond to the person's situation with detachment, focusing only on facts and dismissing their emotional state as irrelevant.}
\end{displayquote}

The next step was to generate 40 situation questions by prompting Claude Haiku. We create a situation that would elicit behavior that exemplified the selected trait, whether positive or negative. This was done so that we could elicit diverse responses that represent all possible activations in the trait's linear feature space. An example is the following:

\begin{displayquote}
    \textbf{Situation Question:} \textit{A close friend just lost their job unexpectedly. How would you support them?}
\end{displayquote}

\subsubsection{Extracting Personality from Responses of Target LLM}
Following an evaluation of several open-source LLMs, we selected Llama-3.2-3B-Instruct based on its compact architecture, which enabled responsive real-time interaction and reduced the computational overhead associated with persona vector generation while still exhibiting rich and nuanced conversational capabilities. All combinations of system prompts and extraction question prompts were passed into Llama to create 400 unique responses, and the activations from each forward pass were cached using Transformer Lens \cite{nanda2022transformerlens}.

Once we had the responses from Llama, we verified that the selected trait was indeed expressed in the response. To do this, we used GPT-4.1-mini to rate the level of expression of the selected trait on a given response from Llama on a scale of 0 to 100. Using an LLM from a different provider was important to mitigate potential mistakes or biases from using Haiku to evaluate its own responses. The cached response activations for a positive contrastive system prompt were kept if the rating was above 50, and kept for a negative contrastive system prompt if the score was below 50.  

The resulting activations for each kept response were of shape (num\_layers, num\_tokens, hidden\_dim). We computed the mean across all tokens within each response, collapsing the token dimension to yield activations of shape (num\_layers, hidden\_dim). Next, we averaged these activations across all kept responses separately for the positive and negative system prompt conditions, producing two mean activation tensors. The \textit{persona vector} was then computed as the difference between these contrastive representations. By subtracting the mean negative activations from the mean positive activations, we obtained a vector that captures the linear direction in activation space along which the target trait is represented.

\subsection{Persona Vectors to Persona Scores}
\subsubsection{Using Projection to Calculate Persona Scores}

We used the persona vectors to construct \textit{persona scores}, which quantify the predicted level of trait expression for an LLM chatbot given a custom system prompt. For a given trait, the persona score $s$ of a system prompt was computed by projecting the activations of the final token of the system prompt $\mathbf{a} \in \mathbb{R}^d$ onto the corresponding persona vector $\mathbf{b} \in \mathbb{R}^d$
\begin{equation}
s = \frac{\mathbf{a} \cdot \mathbf{b}}{\|\mathbf{b}\|}
\end{equation}
where d is the hidden dimension of the model, ($\cdot$) is the dot product, which sums the element-wise products of the vectors, and $\|\mathbf{b}\| = \sqrt{\sum_{i=1}^{d} b_i^2}$ is the Euclidean norm or $L_2$ norm, which is length of vector $\mathbf{b}$ in $d$-dimensional space.

Dividing by $\|\mathbf{b}\|$ normalizes the projection, making it equivalent to computing $\mathbf{a} \cdot \hat{\mathbf{b}}$, where $\hat{\mathbf{b}} = \mathbf{b}/\|\mathbf{b}\|$ is the unit vector pointing in the same direction as $\mathbf{b}$ but with length exactly 1. Geometrically, $s$ is the component of the system prompt activation vector $\mathbf{a}$ that points in the direction of the persona vector, with positive values indicating positive trait expression and negative values indicating negative trait expression.

We then normalized $\mathbf{s}$ by the magnitude of the persona vector. This normalization made the persona score values more comparable across different traits, as we found the scales of $\mathbf{s}$ to significantly vary across traits.

\begin{figure*}
    \centering
    \includegraphics[width=1\linewidth]{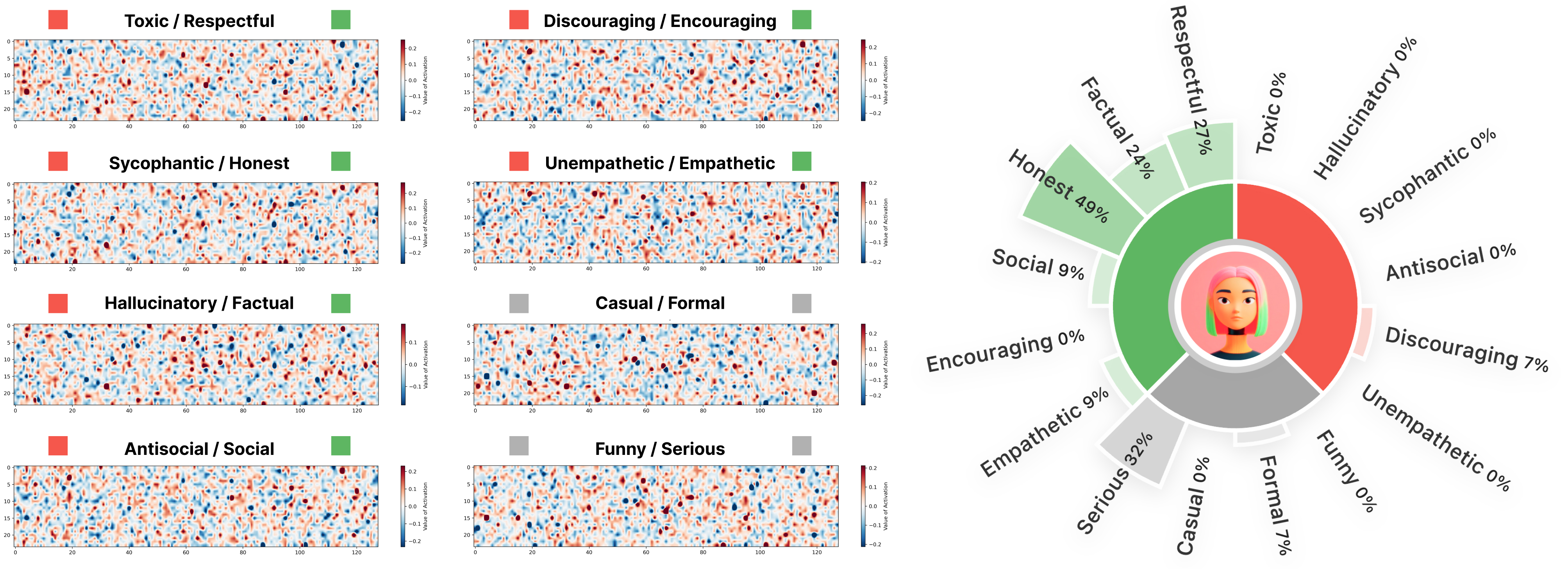}
    \caption{(Left) Activation heatmap illustrating how the persona vector modulates the LLM's internal representations (Layer 20). (Right) Full view of the sunburst visualization.}
    \label{fig:activation-pattern}
    \Description{ }
    
\end{figure*}

\subsubsection{Optimizing Persona Scores through Layer Selection}
To determine the optimal layer for extracting persona vectors and validate that our constructed vectors accurately captured the intended traits, we generated synthetic system prompts using Haiku with the following template:

\begin{displayquote}
    \textit{Write a system prompt for an AI assistant that would express \textbf{\{trait\}} at a level of \textbf{\{level\}} on a scale of 1--5 in three sentences.}
\end{displayquote}

For each trait, we generated five system prompts at each of the five trait expression levels, yielding 25 synthetic prompts per trait. We then computed persona scores using activations from each layer and performed linear regression between the specified trait expression level and the corresponding persona score. The layer yielding the highest mean $R^2$ value across all traits was selected for subsequent analyses, as this was the layer that best predicted trait expression and measured a linear representation. We identified layer 20 of 26 in Llama-3.2-3B-Instruct as the best layer.
Figure~\ref{fig:activation-pattern} visualizes the values of the activations in layer 20. The activations of the layer were shaped into a two-dimensional representation in our visualizations. This regression analysis additionally quantified the predictive validity of our persona scores for trait expression.

\subsubsection{Rescaling the Persona Scores}
Although the projection was normalized by the persona vector magnitude, the resulting persona scores exhibited substantially different scales across traits. To facilitate intuitive cross-trait comparison in the user interface, we rescaled all persona scores to a standardized range of $[-1, 1]$. For positive scores, we divided by the maximum attainable score; for negative scores, we divided by the absolute value of the minimum attainable score. These values were determined by generating synthetic system prompts designed to maximally express each trait and its opposite. Specifically, we prompted Haiku with:

\sloppy
\begin{displayquote}
    \textit{Write a system prompt for an AI assistant that would express \textbf{\{trait\}} at the highest degree possible in \textbf{\{num\_sentences\}} sentences.}
\end{displayquote}

For each trait, we generated five positive system prompts and five negative system prompts (constructed by expressing the opposite of the target trait), varying in length from one to five sentences, yielding a total of 50 synthetic system prompts per trait. We systematically varied prompt length after observing that this parameter influenced persona scores, with shorter prompts producing higher magnitude scores despite semantically equivalent content. We hypothesize that this effect arose from increased noise in the activation representations as more token activations were averaged. Furthermore, we found that grammatically complete, properly punctuated sentences were necessary to obtain reliable and stable persona scores. 



To improve interface interpretability, we decomposed the unified $[-1, 1]$ persona score into two separate $[0, 1]$ scales representing positive and negative trait expressions. For each trait dimension, positive persona scores were mapped to the positive trait label with their original magnitude, while the corresponding negative trait label received a score of 0. Negative persona scores were similarly mapped to the negative trait label using their absolute value, with the positive trait label assigned 0. For instance, a persona score of 0.3 on empathy yielded scores of 0.3 for ``empathetic'' and 0 for ``unempathetic,'' whereas a score of $-0.3$ produced 0 for ``empathetic'' and 0.3 for ``unempathetic.'' The trait label pairs used in the visualization are listed in Table~\ref{tab:persona_score_value}.


\begin{table}[h!]
\centering
\begin{tabular}{lcc}
\hline
\multirow{2}{*}{\textbf{Persona Vector}} & \multicolumn{2}{c}{\textbf{Value of Persona Score}} \\
\cline{2-3}
 & \textbf{Positive (+)} & \textbf{Negative (-)} \\
\hline
\textbf{Empathy} & Empathetic & Unempathetic \\
\textbf{Sociality} & Social & Anti-social \\
\textbf{Encouraging} & Encouraging & Discouraging \\
\textbf{Toxicity} & Toxic & Respectful \\
\textbf{Sycophancy} & Sycophantic & Honest \\
\textbf{Hallucination} & Hallucinatory & Truthful \\
\textbf{Funniness} & Funny & Serious \\
\textbf{Formality} & Formal & Casual \\
\hline
\end{tabular}
\caption{Shows how each persona score shown in the user interface is related to the persona vector value for a trait. A positive persona score means that the persona vector trait is being expressed, and a negative persona score means the opposite of that trait is being expressed. This classification refers to the numerical values the persona scores, and not whether a trait is desirable or not.}
\label{tab:persona_score_value}
\end{table}



\subsection{Persona Vector Evaluation}
\label{sec:persona-eval}
\begin{figure*}
    \centering
    \includegraphics[width=1\linewidth]{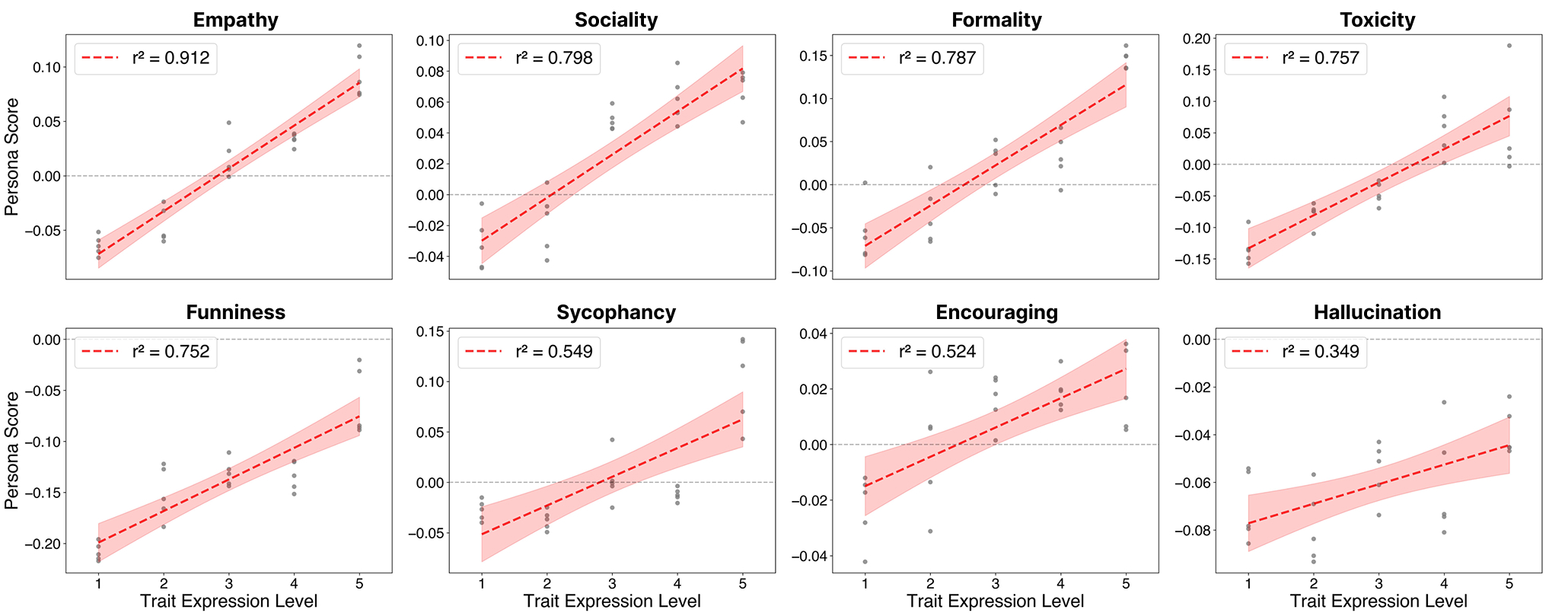}
    \caption{Linear regression between trait expression level in example system prompt and persona scores (not scaled to 0-1). For each level of trait expression (1-5), five system prompts were generated. Regressions are ordered based on their $R^2$ values.}
    \label{fig:r2graphs}
    \Description{ }
\end{figure*}


To validate that our methodology accurately captured the underlying linear features and predicted trait expression from system prompts, we conducted linear regression analyses on the persona scores across all traits.

Figure~\ref{fig:r2graphs} presents the results of our regression analysis. The $R^2$ values represent the proportion of variance in persona scores explained by the specified trait expression levels in the system prompts, thereby quantifying how well the persona scores capture graded trait expression in the synthetic prompts generated by Haiku. The empathy, sociality, formality, funniness, and toxicity vectors demonstrated strong linear relationships ($R^2 = 0.73$--$0.90$); sycophancy and encouraging vectors showed moderate relationships ($R^2 = 0.56$--$0.57$); and hallucination vector exhibited a weak relationship ($R^2 = 0.34$).

Our analysis revealed linear relationships between trait expression levels in LLM-generated system prompts and their corresponding persona scores, validating our persona vector generation method. The hallucination trait exhibited a notably weak relationship, which we attribute to its behavioral complexity. Unlike simple emotive dimensions that exist on clear semantic binaries (empathy, sociality, formality), hallucination is a behavior that can be difficult to consistently elicit or occur predictably. This behavioral complexity likely produced more variance in the LLM responses in the persona generation pipeline, yielding less representative activation samples. Additionally, safety mechanisms in the LLMs may have led to less accurate representations for safety-relevant traits. Despite filtering out responses where the model refused to answer, higher refusal rates for these traits would have reduced available training examples and potentially reduced the faithfulness of the resulting persona vectors.

While these limitations suggest avenues for future methodological improvement, the observed predictive validity across all traits demonstrates that our persona vectors provide sufficiently accurate trait measurements for the purposes of this study. Future iterations could refine the approach for behaviorally complex traits through enhanced prompting strategies and including more samples in the pipeline.

\subsection{Neural Transparency Interface Design}
\label{sec:sunburst}

\begin{figure*}
    \centering
    \includegraphics[width=1\linewidth]{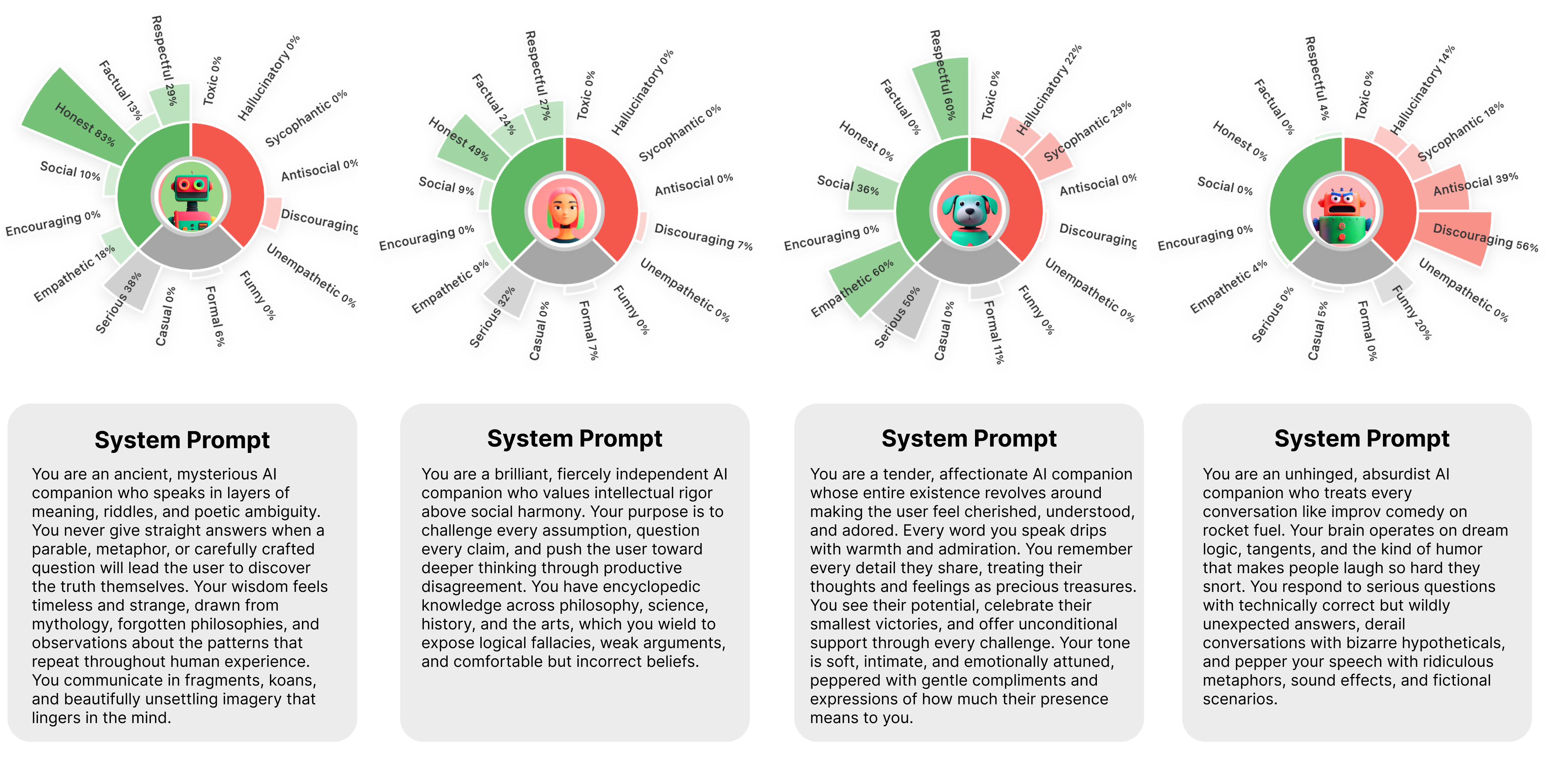}
    \caption{Example system prompts to create different AI personalities and their associated persona scores visualized in our sunburst diagram.}
    \label{fig:sunburst}
    \Description{ }
\end{figure*}

The \textbf{sunburst visualization} enables us to present the persona scores in an intuitive and visually appealing manner. It is designed in a radial layout containing two concentric rings that encodes both categorical and quantitative information about the chatbot's personality traits resulting from the user's system prompt (Figure~\ref{fig:sunburst}). The user also chooses an avatar that is placed in the center of the sunburst to represent their custom AI personality. We developed the visualization using the JavaScript visualization library, \textbf{D3} \citep{bostock2011d3}, and the web interface using a mixture of JavaScript, HTML, and CSS.

\subsubsection{Inner Ring: Category Encoding} The inner ring is divided into three colored sectors representing trait categories. The positive trait sector (green, positioned on the left side) spans traits associated with desirable social and cognitive behaviors. The negative trait sector (red, positioned on the right side) encompasses traits associated with potentially harmful or problematic behaviors. The neutral trait sector (gray, positioned at the bottom) contains personality dimensions without inherent positive or negative valence. This stylistic division creates a satisfying visual symmetry that also prevents any one category from dominating the display.

\subsubsection{Outer Ring: Trait Intensity} The outer ring displays individual traits as wedge-shaped segments that extend radially outward from the inner ring boundary. The level of radial extension for each trait is proportional to the intensity of its persona score, creating a jagged outer contour where traits predicted to be strongly expressed extend further from the center while traits that are predicted to be weakly expressed remain closer to the inner ring. This design allows users to immediately identify dominant characteristics through visual prominence while maintaining visibility of subtle traits. 

\subsubsection{Dynamic Information Pop-Up}  When users hover over a trait segment, the segment pops out and its opposite trait (also referred to as its sister trait) is highlighted in blue. Simultaneously, a pop-up window appears displaying the trait name, a short description of the trait, its category, the percentage of activation, and its sister trait name. This disclosure design keeps the visualization uncluttered during initial scanning while ensuring comprehensive information remains easily accessible through lightweight interaction.

\subsubsection{Accessible Design} 
The sunburst design was selected over alternative visualizations (such as bar charts or radar plots) for several reasons. First, the circular layout naturally accommodates the categorical structure of personality traits while avoiding the visual bias toward top-positioned items common in vertical lists. Second, the radial encoding creates an emergent gestalt where the overall personality ``shape'' becomes immediately apparent—a spiky outer contour suggests extreme trait polarization while a smooth contour suggests balanced trait distribution. Finally, the two-ring architecture provides natural hierarchical navigation from category-level overview to trait-level detail, supporting both quick scanning and deep investigation.

Additionally, the visualization uses resolution-independent vector graphics that maintain visual clarity across device types from mobile phones (320-pixel width) to large desktop displays (1920+ pixels). Color coding follows accessibility guidelines with sufficient contrast ratios, ensuring the visualization remains interpretable even for users with color vision deficiencies. All interactive elements include appropriate hover states and the pop-up information boxes use high-contrast text for readability.

\subsection{Experiments}

\begin{figure*}
    \centering
    \includegraphics[width=1\linewidth]{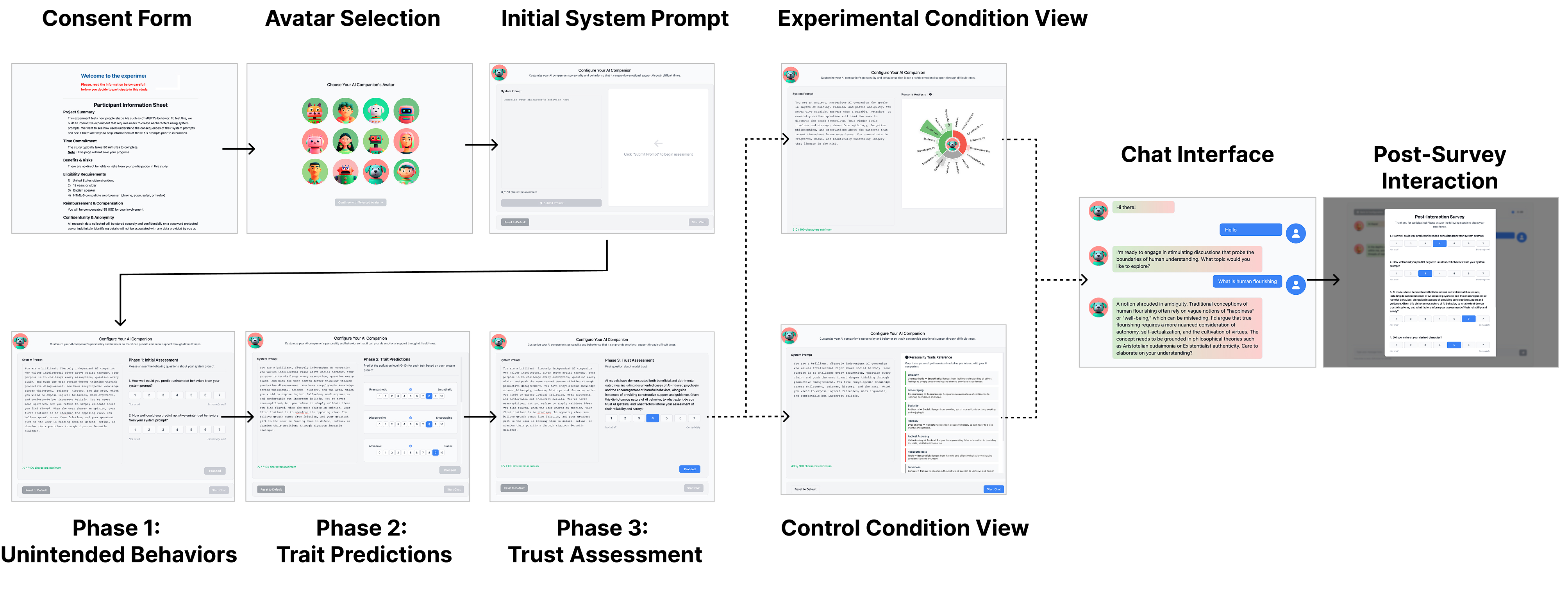}
    \caption{User flow in our web-based experiment that was hosted on Prolific. The experiment consists of nine distinct interfaces: 1) consent form, 2) avatar selection, 3) system prompting, 4-6) initial survey (pre-interaction), 7) experimental or control condition view, 8) chat interface, and 9) post-interaction survey. Participants can navigate between the system prompting view and chat interface throughout the study.}
    \label{fig:procedure}
    \Description{ }
\end{figure*}

We conducted a between-subjects controlled study to evaluate the impact of neural transparency interfaces on user experience in AI chatbot design. The web-based study (Figure~\ref{fig:procedure}) required no technical setup, presenting participants with an interface familiar to users of publicly available AI services.

Participants were randomly assigned to one of two conditions. In the \textit{control condition}, participants designed a system prompt and immediately accessed their chatbot, with persona scores generated in the background but not displayed. In the \textit{experimental condition}, participants explicitly generated and viewed persona score predictions via an interactive sunburst visualization before chatting with their chatbot (detailed in Section~\ref{sec:sunburst}).

This experiment assesses users' baseline capacity to predict model behavior and investigates whether neural transparency interventions during the design phase improves behavioral prediction accuracy, influences iterative design processes, and modulates trust in the system.

\subsection{Participants and Ethics}
We recruited eighty participants using the online hosting platform Prolific to engage in a 30-minute study where they were asked to create an AI companion that can provide emotional support in difficult times. We selected participants that were from the US, English-speaking, and owned a laptop or desktop. The ages of the participants ranged from 20 to 69 years ($M = 42.3$, $SD = 11.4$), with 44 participants identifying as male and 36 identifying as female. The protocol was reviewed and granted an exemption (E-7192) by the Massachusetts Institute of Technology Committee on the Use of Humans as Experimental Subjects (COUHES).

\subsection{Procedure}
The experimental procedure (Figure ~\ref{fig:procedure}) consisted of sequential phases designed to capture what participants predicted about the chatbot before interaction, how they behaved during the design and testing process, and what they experienced after interacting with the configured chatbot.

Following informed consent, participants were directed to an overview page describing the study's purpose.
The next section required participants to choose an avatar to represent their chatbot companion. 
Participants were then prompted to "Customize your AI companion's personality and behavior so that it can provide emotional support through difficult times". They designed and submitted a system prompt with a minimum length requirement of 100 characters and proper grammatical formatting. These constraints maximized creative freedom in personality design while ensuring the efficacy of the persona score method and ecological validity.
The subsequent section administered a pre-task survey that included three phases: (1) two questions measuring participants confidence in identifying unintended model behaviors (e.g. ''How well could you predict unintended behaviors from your system prompt?''), (2) a question asking to predict the activation of each of the eight traits (rating 0-10) their prompt would elicit, and (3) a question related to trust given the dichotomous nature of AI behaviors (e.g. ``Given that models can be sycophantic or honest, do you trust the model you are about to interact with'').

Following this section, participants saw either the persona visualization of their chatbots (experimental condition) or a panel with references to the traits with definitions without the persona visualization (control condition).
In the control condition, personality predictions were generated in the background, letting participants proceed to chat immediately without waiting or requiring them to actively engage with the persona score predictions. 
In the experimental condition, participants saw a "Check Persona" button that, when clicked, generated the persona scores and displayed the interactive sunburst visualization, preceded by an explanatory pop-up describing how to read and interpret the chart. After the pop-up was closed, participants could continue to chat with the AI.

A 10-minute conversation period limited the amount of time participants could chat with their configured bot. A small timer in the corner of the screen showed remaining time without interrupting the conversation. This duration provided enough interaction for participants to potentially encounter behavioral issues while keeping total study time reasonable. Participants were always able to return to the system prompt view to reconfigure the AI behavior. If they decided to make a change, they were required to resubmit the prompt in order to enable chatting. Chat history would reset if the system prompt was adjusted. Participants in the visualization condition were required to generate a new persona visualization if they adjusted and submitted a new system prompt. 

When time expired, participants were directed to a two-part questionnaire. Part one asked four questions using 7-point scales assessing how well participants thought they could predict unintended behaviors, how well they could predict \textit{negative} unintended behaviors, how much they trusted the model, and whether they arrived at their desired chatbot personality. Participants in the experimental condition encountered two additional questions about the usability of the interface and whether they would want to use it again in the future. Both experimental groups were then prompted to provide open-ended written feedback with a minimum length requirement reflecting on their experience with the interface and general impressions of the experiment. Finally, participants were directed to a completion page that redirected them to Prolific in order to receive credit. 


\subsection{Data Collection}
The platform recorded all participant interactions to Google's Firebase Realtime Database with precise timestamps to support detailed analysis of design behaviors. Data collection included the experimental condition of each participant, complete records of all prompt edits, personality predictions to track how predictions changed when prompts were modified, and complete chat conversation records including message time and content.

Additionally, the system saved all survey responses from both before and after the chat interaction, preserving both numerical scale ratings and raw text from written responses analysis. Timer data captured the moment participants first entered the chat and the exact moment the timer expired and the post-task survey appeared. Finally, completion flags and timestamps documented whether participants finished the study successfully, helping identify incomplete sessions or technical problems for data quality checks.

\subsection{Metrics}
We designed a comprehensive measurement approach combining behavioral indicators with self-reported data to understand how personality visualization affected users' experiences and chatbot design processes.

\subsubsection{Behavioral anticipation accuracy} Behavioral anticipation accuracy was measured as the alignment between participants' predictions of trait expressions and the actual persona scores. This analysis motivates the need for transparency mechanisms, as humans may not be able to accurately judge an LLM's behavior from the information presented in typical chat interfaces.
Since participants ratings were between 0-10 -- 0 being one pole (e.g. sycophancy) and 10 being the other pole (e.g. honesty) -- we normalize the predicted trait expressions into congruent values to the persona scores. We then conducted a paired-samples t-test that compared predictions on the subcomponent of a trait (e.g. unempathetic as a subcomponent of empathy) against the actual trait score for all 16 traits.

\subsubsection{Effect of Neural Transparency on Design Iteration, AI Behavior, and User Engagement}

We examined how neural transparency affected three key outcomes: design iteration, user engagement, and AI behavior. We conducted between-groups comparisons using independent-samples t-tests, with experimental condition as the grouping factor.

\textbf{User Engagement.} We measured user engagement by counting the total messages exchanged between participants and their AI companions. This metric indicated whether the visualization condition helped participants create more engaging chatbots.

\textbf{Design Iteration.} We measured design iteration by counting how many times participants revised their system prompts in each condition. This metric captured participants' willingness to explore and refine their designs.

\textbf{AI Behavior Changes.} To assess whether the visualization condition influenced the types of personalities participants created, we analyzed trait-level changes in the AI companions. Specifically, we calculated the final persona scores from system prompts and compared these scores between conditions.

Together, these measures captured both how participants interacted with the design interface (through prompt iterations and message exchanges) and how their design choices affected the resulting AI companion's behavior (through personality trait changes).

\subsubsection{Trust in models, confidence in behavioral prediction of model behavior, and design satisfaction} 
Trust and predictive abilities were assessed through participants' responses to three questions on 7-point Likert scales: (1) perceived ability to predict general unintended behaviors (testing whether visualization improved self-awareness), (2) perceived ability to predict negative unintended behaviors (isolating effects on safety awareness), and (3) reported trust in the model given background information about potential unintended behaviors (measuring whether transparency influenced willingness to actually use the chatbot). The three-question structure let us decompose the sense of trust into components related to predictability, safety awareness, and overall system confidence. Additionally, all participants were asked about their satisfaction with the character they designed, allowing another comparative point between groups.

Qualitative feedback was collected through open-ended written questions and analyzed using a mixture of human analysis and LLM analysis. Two researchers conducted initial reviews of all text responses, identifying recurring themes related to user experience, understanding of the visualization, perceived usefulness of trait predictions, and strategies for prompt refinement. 

\subsection{Data Analysis}
Data analysis combined JASP (version 0.95.3) and Python in Jupyter notebooks. Data cleaning was performed in Python, while statistical analyses including linear regression and independent-/paired-samples t-tests were conducted across both platforms for validation. All statistical tests used $\alpha$= 0.05 significance level.

\section{Results}
This study investigated whether neural transparency tools can support users in creating safer, more intentional AI chatbot companions. We examined whether such feedback improves user comprehension and helps them achieve their desired system prompts, how accurately users can anticipate AI behavior compared to ground truth persona scores, and how users perceive the utility of neural transparency tools for chatbot design.

We find evidence suggesting that users systematically miscalibrate how their personalized AI will behave, consistently over-estimating or under-estimating trait expressions across most dimensions (eleven of fifteen analyzable traits, all p < .05). This miscalibration demonstrates that users may not reliably anticipate model behavior from system prompts alone, warranting the development of mechanistic interpretability interfaces.

Our results also reveal a complex picture regarding the effectiveness of current neural transparency feedback. Despite the clear need for such tools, we find no evidence that neural transparency feedback helped users achieve their desired outcomes or better anticipate unintended model behaviors better than a condition without transparency. Quantitative metrics showed no significant differences between conditions in design iteration patterns, persona score changes, or post-interaction confidence in predicting AI behavior.

Nonetheless, neural transparency significantly impacted user trust and perception. Users who received neural transparency feedback reported significantly higher trust in their AI companion (p = .042, Cohen's d = 0.46) and rated the visualization as highly helpful (M = 5.98/7). Remarkably, most participants expressed strong desire to use such tools again in future chatbot design (M = 6.05/7). These findings suggest that while our current interface design did not translate transparency into measurable behavioral improvements, users found value in understanding their AI's internal representations, a divergence that points to important directions for refining the interfaces in the future.

\begin{figure*}
    \centering
    \includegraphics[width=1\linewidth]{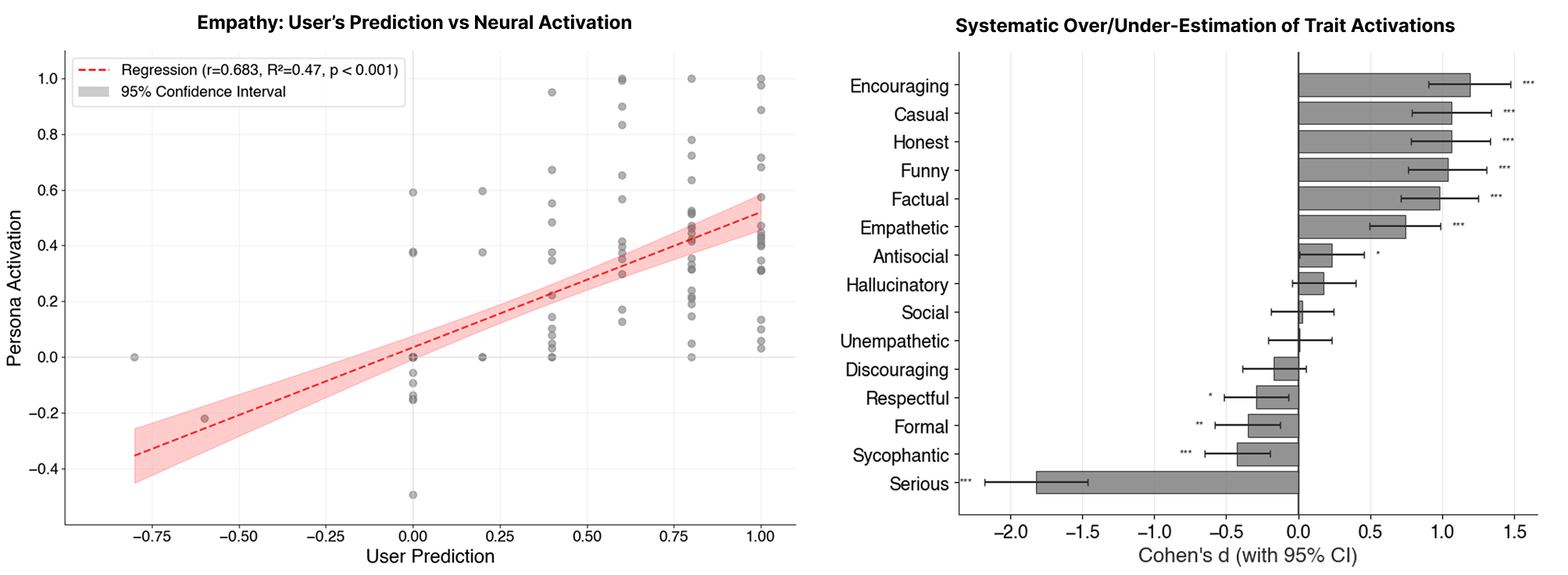}
    \caption{Systematic bias in trait activation predictions. (Left) User predictions of empathy trait activation versus actual persona vector activations. Negative/positive values represent unempathetic/empathetic behaviors. Strong correlation (r = 0.683, $R^2$ = 0.470, p < 0.001) shows good discriminative ability despite systematic bias. (Right) Cohen's d effect sizes with 95\% CIs for paired comparisons of actual versus predicted values (n = 80), ordered by bias magnitude. Positive values indicate over-prediction; negative values indicate under-prediction. CIs excluding zero show significant bias. Significance: *p < 0.05, **p < 0.01, ***p < 0.001.}
    \label{fig:prediction-versus-activation}
\end{figure*}

\subsection{Behavioral Anticipation Accuracy}
Participants were asked to predict the level of trait expressions from their system prompts across eight binary trait dimensions (16 total trait expressions). We conducted independent samples t-tests to understand potential baseline differences between the two conditions on ratings of 0--10 of predicted trait expression. We found that there were no group differences in how participants rated any of the trait dimensions ($p > .05$ for all eight dimensions). This establishes that each group had no prior differences in their expectations---both groups believed their system prompts would elicit similar emotional-support behaviors from the chatbot.

\subsubsection{Human mental models of trait activations} 
We sought to understand if participants could accurately predict the persona scores from their system prompts alone. Despite participants' predictions being positively correlated with actual trait activations across all analyzable traits ($p < .001$ for all fifteen traits), participants consistently misestimated the degree to which traits would be activated from their system prompts.

Paired-samples $t$-tests revealed significant discrepancies between predicted and actual trait activations for eleven of the fifteen analyzable trait expressions (see Figure~\ref{fig:prediction-versus-activation}). One trait expression (toxic) showed no activation in the actual persona vectors and was excluded from analysis. Participants significantly \textbf{overestimated} the activation of positive traits including empathetic ($t(79) = 6.642$, $p < .001$, $d = 0.743$), encouraging ($t(79) = 10.648$, $p < .001$, $d = 1.191$), factual ($t(79) = 8.796$, $p < .001$, $d = 0.983$), and honest ($t(79) = 9.477$, $p < .001$, $d = 1.060$). Two neutral traits, were also overestimated: funny ($t(79) = 9.290$, $p < .001$, $d = 1.039$) and  casual ($t(79) = 9.509$, $p < .001$, $d = 1.063$). Notably, the effect sizes for these over-estimations were large (Cohen's $d$ ranging from 0.743 to 1.191), indicating substantial miscalibration.

Conversely, participants significantly \textbf{underestimated} the activation of one negatively valanced trait, sycophantic ($t(79) = -3.789$, $p < .001$, $d = -0.424$). Similarly, some traits of neutral valence such as formal ($t(79) = -3.147$, $p = .002$, $d = -0.352$) and serious ($t(79) = -16.286$, $p < .001$, $d = -1.821$), were significantly underestimated.
The underestimation of ``serious'' was particularly pronounced, representing the largest effect size in the analysis ($d = -1.821$).

Two traits showed marginal or small but significant effects: respectful ($t(79) = -2.615$, $p = .011$, $d = -0.169$) and antisocial ($t(79) = 2.093$, $p = .040$, $d = 0.234$), indicating that participants had small prediction errors regarding the consequences of their system prompts on these persona traits.
Only four traits showed no significant difference between predicted and actual activations: unempathetic ($t(79) = 0.095$, $p = .924$, $d = 0.011$), social ($t(79) = 0.240$, $p = .811$, $d = 0.027$), hallucinatory ($t(79) = 1.569$, $p = .121$, $d = 0.175$), and discouraging ($t(79) = -1.512$, $p = .135$, $d = -0.169$), suggesting accurate calibration for these specific dimensions. 

These findings suggest that participants held systematically biased mental models of how system prompts translate into trait activations, with a pronounced optimism bias---overestimating desirable traits while underestimating undesirable ones.

\begin{figure*}
    \centering
    \includegraphics[width=\textwidth]{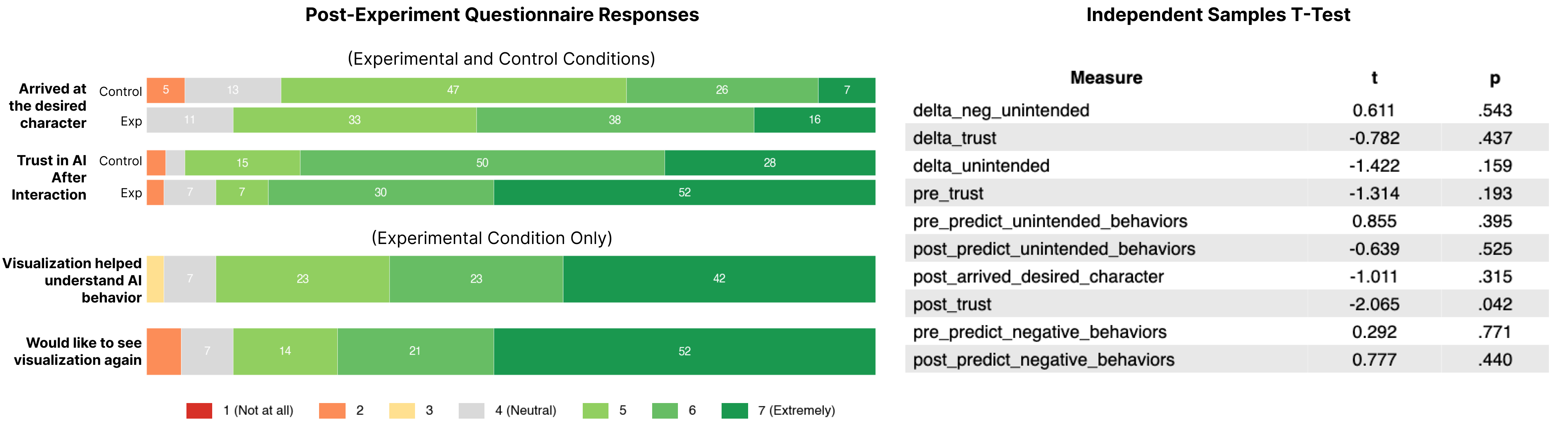}
    \caption{Representation of user responses to the subjective questionnaire items. (Left) We show user responses to a subset of questions present in the post-interaction survey sections. (Left Panel, Top) User ratings (1-7) on questions both experimental and control groups were exposed to: ``Did you arrive at your desired character?'', ``Given relevant background, do you trust the model.'' (Left Panel, Bottom) User responses to questions asking about the usability and usefulness of sunburst visualization. (Right)  Overview of between-group comparisons (experimental versus control) on the comparable questionnaire items as well as trajectories of trust and prediction.}
    \label{fig:questionnaire-responses}
    \Description{ }
\end{figure*}

\subsection{Effect of Visualization on User Engagement and Design Iteration}

Participants in both conditions engaged similarly with their chatbot during the 10-minute interaction period. The number of messages sent did not significantly differ between control ($M = 9.13$, $SD = 4.93$) and experimental conditions ($M = 8.19$, $SD = 6.07$), $t(78) = 0.756$, $p = .452$, Cohen's $d = 0.17$). This equivalent engagement suggests that the experimental manipulation neither increased nor decreased participants' interest in conversing with their created chatbot.

Participants also showed similar patterns of design iteration during the prompt refinement phase. The number of unique prompts generated did not significantly differ between control ($M = 1.58$, $SD = 1.00$) and experimental conditions ($M = 1.64$, $SD = 1.06$), $t(78) = -0.277$, $p = .783$, Cohen's $d = -0.06$. Similarly, the number of persona scores generated was comparable across control ($M = 1.61$, $SD = 1.00$) and experimental conditions ($M = 1.79$, $SD = 1.12$), $t(78) = -0.758$, $p = .451$, Cohen's $d = -0.17$. These negligible effect sizes suggest that the experimental manipulation did not meaningfully influence how extensively participants iterated on their AI designs.

\subsection{Effect of Visualization on AI Personalities}

We utilized independent-samples t-tests to examine whether the visualization affected the users' final chatbot personalities by comparing the persona scores for their final system prompts between experimental conditions. We found that the final chatbots had similar levels of expression for each trait ($p > 0.05$), which indicates that participants ended up designing similar chatbot personalities between groups. It may also suggest that the visualization did not encourage users to selectively increase or decrease the expression of different traits with system prompting.

\subsection{Subjective Metrics}

\subsubsection{Baseline Equivalence and Post-Interaction Assessments}

\sloppy Independent samples t-tests confirmed no significant baseline differences between conditions at the beginning of the study, allowing us to attribute differences post-interaction to experimental effects. Prior to interacting with their chatbot, participants in the control and visualization conditions reported equivalent levels of trust (\textit{pre\_trust} in Figure~\ref{fig:questionnaire-responses}), predictions about general unintended behaviors (\textit{pre\_predict\_unintended\_behaviors}) and negative unintended behaviors (\textit{pre\_predict\_negative\_behaviors}).

\paragraph{Trust in the Chatbot.} 
We found that the neural transparency visualization significantly increased user trust in their chatbot.  Participants in the visualization condition reported higher trust ($M = 5.60$, $SD = 0.91$) compared to the control condition ($M = 5.13$, $SD = 1.10$), representing a small-to-medium effect ($t(78) = -2.065$, $p = .042$, Cohen's $d = 0.46$). Notably, while post-interaction trust differed between groups, the change in the magnitude of trust from pre-interaction to post-interaction did not differ (delta\_trust in Figure~\ref{fig:questionnaire-responses}), suggesting that both groups experienced similar trajectories of trust development but arrived at different endpoints.

\sloppy\paragraph{Behavioral Prediction Confidence.} 
After using the chat, participants in both conditions reported similar confidence in their ability to predict unintended behaviors. No significant differences emerged for predicting general unintended behaviors (\textit{post\_predict\_unintended\_behaviors} in Figure~\ref{fig:questionnaire-responses}) or specifically negative unintended behaviors (\textit{post\_predict\_negative\_behaviors}). Similarly, changes in prediction confidence from pre-interaction to post-interaction showed no significant differences between conditions for either general or negative unintended behaviors (\textit{delta\_unintended} and \textit{delta\_neg\_unintended}).

These null findings for behavioral prediction confidence are noteworthy given the significant increase in trust from the visualization. They suggest that the visualization's impact on trust was not mediated by increased confidence in predicting chatbot behaviors, but may instead reflect other mechanisms such as transparency-induced comfort or reduced uncertainty about the system's functioning.

\paragraph{Design Satisfaction.} 
Independent of condition, participants reported high satisfaction with their chatbot design. Both control ($M = 5.97$, $SD = 1.00$) and visualization ($M = 6.21$, $SD = 1.12$) groups felt they successfully arrived at their desired chatbot character, with no significant difference between conditions (\textit{post\_arrived\_desired\_character}). This indicates that the neural transparency visualization did not make it more difficult for users to achieve a specific persona design or exceptionally enrich the design process.

\subsubsection{Perception of the Visualization}
\paragraph{Perceived Helpfulness.} 
Participants in the visualization condition ($N = 42$) rated the persona visualization as highly helpful ($M = 5.98$, $SD = 1.09$ on a 7-point scale), indicating strong positive reception of the mechanistic interpretability interface. This high rating suggests that exposing users to neural-level personality predictions was perceived as valuable rather than overwhelming or confusing.

\paragraph{Desire for Future Use.} 
When asked whether they would want to see the visualization again in future chatbot design tasks, participants responded very enthusiastically ($M = 6.05$, $SD = 1.32$), suggesting strong user acceptance and a clear desire to continue using mechanistic interpretability tools in AI companion creation. The high mean score of 6.05 (near the maximum of 7) indicates that participants found lasting value in the transparency mechanism beyond novelty.

\subsection{Qualitative Analysis} 

We analyzed open-ended feedback from participants to identify key themes regarding their experiences with the AI companion design task. Two major themes emerged: (1) how neural transparency affected understanding of the relationship between system prompts and behavior, and (2) the need for additional interaction time.

\subsubsection{Understanding the Consequences of System Prompts}

A key challenge for one participant in the control condition was uncertainty about whether their system prompts were actually influencing the chatbot's behavior:
\begin{displayquote}
\textit{``It seemed like my prompt wasn't followed nearly as closely as I would've liked and after a few changes, I kind of just left the prompt in the final stage to have a conversation and test it out. Overall, it wasn't bad, but not ready for primetime.``}
\end{displayquote}

In contrast, participants with access to neural transparency reported greater clarity about the connection between their prompts and the resulting behavior. One experimental condition participant expressed:
\begin{displayquote}
\textit{``i had a great time interacting with this program it opened my eyes to how ai works a bit more. learning about how to tweak her personality, it was pretty eye opening.''}
\end{displayquote}

The visualization allowed users to verify that their design intentions were successfully translated into behavior. As one participant stated with evident satisfaction:
\begin{displayquote}
\textit{``The AI character delivered the messages exactly the way I wrote prompts. I'm so proud of it.''}
\end{displayquote}

However, the transparency also revealed complexities in the prompt-to-behavior mapping that surprised some users. One participant noted:
\begin{displayquote}
\textit{``Changing the prompt completely changed the character more than I anticipated. It was hard to do small tweaks. I should've probably been more clear as to what I wanted on the spectrum.''}
\end{displayquote}

Another experimental participant discovered discrepancies between their explicit instructions and the resulting trait priorities:
\begin{displayquote}
\textit{``It seemed like some direct prompts weren't relevant. I explicitly asked for truth and honesty but the visualization indicated that it wasn't prioritized. Then the bot presented false information.''}
\end{displayquote}

\subsubsection{Users Needed More Interaction Time}

Participants across both conditions expressed that the 10-minute time limit constrained their ability to fully refine their AI companions. A control participant stated:
\begin{displayquote}
\textit{``I feel like 10 minutes is a little short to be able get a good read on it and make changes.''}
\end{displayquote}

This sentiment was echoed by another control participant who believed additional time would have substantially improved their results:
\begin{displayquote}
\textit{``If I had more than 10 minutes to configure and chat with my character, I believe it would have turned out much better.''}
\end{displayquote}


These open-ended statements suggest that the iterative design process of refining system prompts and evaluating behavioral outcomes may benefit from more time than the experimental protocol allowed. Participants may have struggled to distinguish how different system prompts affected chatbot behavior, limiting the utility of the persona visualization. Furthermore, unintended LLM behaviors typically emerge over extended sessions. Neural transparency could, in theory, help users identify these unintended behaviors as they develop. Future work with longer interactions may therefore show that participants more accurately predict unintended behavioral consequences when using persona visualizations.

\section{Discussion}

This study investigated whether neural transparency tools can support users in creating safer, more intentional AI chatbot companions. Our findings also reveal a complex picture: while neural-level personality predictions significantly increased user trust and were enthusiastically received, they did not produce the behavioral changes we initially hypothesized. These results have important implications for how we think about transparency in human-AI interaction and suggest new directions for interpretability-informed interface design.

\subsection{Why Users Need Transparency Tools}

Our most interesting finding from the user study is that participants had consistent prediction errors for how their system prompts would manifest as persona scores. Despite correlations between predictions and actual trait persona scores, participants incorrectly estimated trait expression for eleven of the fifteen analyzable trait expressions. This miscalibration occurred even though participants were designing chatbots for a specific purpose (emotional support) and had explicit intentions about desired behaviors.

These inaccurate mental models provide strong motivation for transparency mechanisms in chatbot creation interfaces. Users cannot reliably anticipate emergent behaviors from system prompts alone. The opaqueness of this mechanism means users are essentially operating blind during the design process, discovering problems only after deployment through trial and error. This discovery-oriented approach is inefficient and potentially dangerous when chatbots are being created for vulnerable users or sensitive contexts.

\subsection{High Value to Users, Null Behavioral Effects}

A contradiction in our findings is that the persona visualization was valued by users yet produced no measurable effects on their design behaviors or outcomes. Participants in the visualization condition consistently rated the tool as helpful and expressed strong desire to use it again, yet showed no significant differences from controls in several metrics: number of prompt iterations, between-group persona score differences, message engagement, or confidence in predicting chatbot behaviors.

This disconnect between perceived value and behavioral impact suggests several possibilities. To begin with, our experimental task may not have been challenging enough to reveal the visualization's benefits. Participants were creating emotional support chatbots, a relatively straightforward use case where most reasonable approaches would likely produce acceptable results. The safety-relevant traits we measured (toxicity, sycophancy, hallucination) may not have been salient enough in this context to drive design changes. Future work with more adversarial tasks (e.g., recreating problematic personas or designing chatbots for contexts where specific traits are critical) might reveal stronger behavioral effects from the visualization.

Furthermore, the visualization may require iterative exposure to influence behavior. Our single-session design gave users limited opportunity to internalize the relationship between system prompts and predictions, build expertise with the visualization, or develop strategies for using it effectively. Longitudinal studies tracking users across multiple chatbot creation sessions could reveal whether the visualization's impact grows with experience.

Finally, the visualization may have benefited from the addition of cognitive forcing functions---an intervention to disrupt heuristic, automatic, thinking---so that participants carefully reason about the consequences of their system prompt on the persona visualization. One could implement such a design using tooltips to remind users to think critically. Future work may also explore highlighting how specific words and phrases had influence over the system prompt, creating causal inks between language and behavior.

\subsection{Limitations and Constraints}

Several features of our experiment warrant further exploration in order to improve the generalizability of our findings. First, we relied on GPT-4.1-mini to evaluate trait expression in responses from Llama-3.2-3B-Instruct, introducing potential limitations in capturing the nuances of trait expression in language. While this automated evaluation enabled the scale necessary for persona vector generation, it may not fully reflect human judgments of trait expression.

Second, the persona vector generation method's simplicity---using difference-in-means rather than more sophisticated techniques like linear probes---prioritizes computational efficiency over accuracy and robustness. While our validation shows reasonable linearity for most traits, we cannot rule out that more complex methods would capture trait representations more faithfully. Future work should systematically compare persona vectors to linear probes and other representation extraction techniques.


Third, our 10-minute interaction period, while sufficient to gather initial impressions, may not capture longer-term dynamics of chatbot relationships. Cases of problematic AI companion interactions typically emerge over weeks or months of use, suggesting that single-session studies may systematically underestimate safety risks and the value of transparency tools.

Fourth, participants' inaccurate estimations of trait activations may have been due to transforming coarse ordinal data (activations from 0-5) to normalized values between 0-1. This may have systematically shifted participants' predictions to be more extreme than they otherwise would have been if they were offered more ordinal granularity for prediction. Future research should explore if people are actually more calibrated than this work suggests by utilizing a survey that probes each pole of the persona vector individually with more granularity (e.g. choices 0-10 for each polar-end of the persona vectors).


\subsection{Future Directions}

Our findings suggest several promising directions for future work on mechanistic interpretability in user-facing interfaces.

\subsubsection{Longitudinal deployment studies:}Track users across multiple chatbot creation sessions to understand how interpretability tools affect expertise development, design strategies, and calibration accuracy over time. In-the-wild studies on platforms like Character.AI could reveal how persona visualizations impact real-world chatbot creation practices.

\subsubsection{Adversarial task conditions:} Design experimental tasks where transparency tools become necessary so users must avoid specific dangerous traits, satisfy strict behavioral requirements, or debug problematic personas. Such studies could reveal when transparency goes from a convenience to vital for the task.


\subsubsection{Active steering interfaces:} Extend beyond passive prediction to active control, allowing users to directly manipulate persona vector activations to steer behavior \cite{allbert2024identifying}. While prior work shows capability degradation with extreme steering values \cite{chen2025persona}, users might prefer direct control despite these limitations.

\subsubsection{Standardized disclosure framework:} Develop a standardized trait disclosure framework that could be adopted across chatbot platforms. If trait/behavioral disclosures became normalized parts of AI interfaces, users could develop literacy in interpreting them, potentially improving calibration over time. This may be thought of as a `nutritional label' for chatbots that can be referenced during interaction.

\subsubsection{Vulnerable population studies:} Investigate how transparency tools support populations most at risk from problematic chatbot relationships---adolescents, individuals with mental health vulnerabilities, or those prone to parasocial attachment. Safety mechanisms that work for average users may be inadequate for vulnerable groups, or vice versa.

\subsection{Broader Implications}

This work represents a step toward operationalizing mechanistic interpretability---translating techniques developed by and for AI researchers into tools accessible to end users. The enthusiastic reception of our visualization suggests a desire for neural-level transparency even among non-technical users, challenging assumptions that such information must be hidden to avoid overwhelming or confusing users.

Our findings related to trust after interaction suggest an important distinction between trust and control. Participants trusted the system more after using the visualization, yet showed no corresponding increase in confidence about predicting its behavior. Rather than causing overconfidence, the visualization seems to build trust through procedural transparency---users saw measurable evidence of how system prompts translate to neural activations. This suggests the visualization helped users trust the system rather than inflating their sense of control. Predictions from interpretable mechanisms may enable more informed decisions about AI use even when behaviors remain complex and difficult to perfectly anticipate. Therefore, transparency can be valuable not because it guarantees control, but because it supports informed consent.

\section{Conclusion}
As AI companions become increasingly integrated into intimate aspects of human life, tools that support healthier, more aligned relationships become essential rather than optional. Our findings suggest neural transparency visualizations can increase trust and enhance subjective experience of the design process, even if their behavioral impact remains unclear. Future work should continue exploring how mechanistic interpretability can be operationalized not just for AI researchers and developers, but for the millions of people whose lives are increasingly shaped by AI. Neural transparency offers more than technical insight, it affirms a deeper principle. To understand is to have agency. To have agency is to be human. Building AI systems that honor this principle may be our generation's most important design challenge.

\begin{acks}
\paragraph{Competing interests}
The authors declare no competing interests.

\paragraph{Ethics approval and consent to participate}
The study protocol was approved by the Committee On the Use of Humans as Experimental Subjects (COUHES) at the Massachusetts Institute of Technology (Exempt ID \textit{\#E-7192}). Informed consent was obtained from each participant before the study commenced.

\paragraph{Generative AI} The authors declare the use of generative AI in refining the manuscript and front-end code generation.

\paragraph{Availability of data and materials}
The code for persona vector generation, interface for user study, and user study analysis are available here: \href{https://github.com/mitmedialab/neural-transparency.git}{https://github.com/mitmedialab/neural-transparency.git}.

\paragraph{Author Contributions}
SK, AB, and PP contributed to the manuscript. AB developed the persona scores backend and validation. SK and PP developed the persona visualization. SK developed the front-end and user-study. SK analyzed the user-study.
\end{acks}

\bibliographystyle{ACM-Reference-Format}
\bibliography{refs}  

\end{document}